\documentclass[12pt,twoside]{article}

\evensidemargin=\oddsidemargin
\def\Title#1#2#3{%
    \baselineskip=18pt
    \begin{center}
          {\large\bf\uppercase{#1} \\ }
          \bigskip\bigskip
          {#2} \\
          {#3} \\
    \end{center}}
\long\def\Abstract#1{%
         \bigskip
         \parbox{0.93\textwidth}{%
                 \begin{center}
                       {\bf Abstract} \\
                 \end{center}
                 \medskip{\baselineskip=14pt #1}
                 \vss}
         \bigskip}

\makeatletter
\renewcommand{\section}%
 {\@startsection{section}{1}{0pt}%
  {-3.25ex plus -1ex minus -.2ex}{1.5ex plus .2ex}%
  {\vspace*{5mm}\raggedright\large\bf }}
\renewcommand{\thesection}{\arabic{section}.}
\@addtoreset{equation}{section}
\renewcommand{\@eqnnum}{(\thesection\theequation)}
\renewcommand{\p@equation}{\thesection}
\makeatother

\begin{document}

\vspace*{1cm}

\Title{COSMOLOGICAL SOLUTIONS FOR THE UNIVERSE \\
FILLED WITH MATTER IN VARIOUS STATES \\
AND GAUGE INVARIANCE}%
{T. P. Shestakova}%
{Department of Theoretical and Computational Physics, Rostov State University, \\
Sorge St. 5, Rostov-on-Don 344090, Russia \\
E-mail: {\tt shestakova@phys.rsu.ru}}

\Abstract{We explore at phenomenological level a model of the Universe
filled with various kinds of matter characterized by different
equations of state. We show that introducing of each kind of matter
is equivalent to a certain choice of a gauge condition, the gauge
condition describing a medium with a given equation of state. The
case of a particular interest is when one kind of matter (de
Sitter false vacuum) dominates at the early stage of the Universe
evolution while another kind (radiation, or ultrarelativistic gas)
dominates at its later stage. We can, therefore, consider different
asymptotic regimes for the early and later stages of the Universe
existence. These regimes are described by solutions to the Wheeler
-- DeWitt equation for the Universe with matter in that given
state, and, at the same time, in the ``extended phase space''
approach to quantum geometrodynamics the regimes are described by
solutions to a Schr\"odinger equation associated with a choice
of some gauge condition. It is supposed that, from the viewpoint of
the observer located at the later stage of the Universe evolution,
solutions for a $\Lambda$-dominated early Universe would decay.}

\section{Introduction}
It has been realized in recent years that the main problems of the
Wheeler -- DeWitt quantum geometrodynamics, such as the famous
problem of time and a related problem of Hilbert space, cannot be
solved without fixing a reference frame. So, in \cite{BK} a
privileged reference frame was fixed by introducing an incoherent
dust which plays the rule of ``a standard of space and time''. In
\cite{MM} the so-called kinematical action was introduced
describing a dust reference fluid. Fixing a reference frame
results in replacing the Wheeler -- DeWitt equation
\begin{equation}
\label{WDW-eq}
H\Psi=0
\end{equation}
by a Schr\"odinger equation
\begin{equation}
\label{Sch-eq}
i\frac{\partial\Psi}{\partial t}=H\Psi,\quad
\rm{or}\quad
H\Psi=E\Psi.
\end{equation}
(see also the recent paper \cite{BM}, where the authors start from
Schr\"odinger, or evolutionary quantum gravity; for a review on the
problem of time and gauge invariance in quantum cosmology see
\cite{SS1,SS2}).

The same purpose (the replacement of the Wheeler -- DeWitt static
picture of the world by a Schr\"odinger evolutionary dynamics) was
achieved in an explicitly gauge noninvariant approach named
``quantum geometrodynamics in extended phase space''
\cite{SSV1,SSV2,SSV3,SSV4}. In these works it was demonstrated
that, firstly, quantum theory of the Universe with non-trivial
topology cannot be constructed in a gauge invariant way, and,
secondly, the introduction of a gauge fixing term into the
gravitational action can be interpreted as the existence of some
media with a certain equation of state in the Universe.

On the other side, even in the limits of canonical quantization of
gravity the Wheeler -- DeWitt equation can be reduces to a
Schr\"odinger-like equation $\tilde H\Psi=E\Psi$, where $\tilde
H-E=0$ is a new form of the Hamiltonian constraint, and $E$ is a
conserved quantity, which appears from phenomenological
consideration of matter fields with various equations of state,
including de Sitter false vacuum. Solving this equation, one would
find an energy spectrum for this sort of matter. This approach was
discussed, in particular, in the works by Dymnikova and Fil'chenkov
\cite{Fil,DF}. The purpose of this talk is to compare the both
approaches and show that, in some sense, introducing of each kind
of matter is equivalent to a certain choice of a gauge condition.

\section{Canonical quantization approach}
To give a qualitative discussion of the question, I shall advisedly
restrict myself to the simplest isotropic closed model with the
action
\begin{equation}
\label{action-1}
S=-\!\int\!dt\,\left(\frac12\frac{a\dot a^2}N
  -\frac12Na+Na^3\Lambda\right)+S_{(mat)},\quad
S_{(mat)}=-\!\int\!dt\,Na^3\varepsilon(a).
\end{equation}
Matter fields are presented in this model phenomenologically,
without a clear indication on the nature of the fields, by its
energy density $\varepsilon(a)$, the dependence of $\varepsilon$ on
$a$ determining its equation of state. It is easy to check that for
\begin{equation}
\label{eps-n}
\varepsilon(a)=\frac{\varepsilon_n}{a^n}
\end{equation}
$\varepsilon_n={\rm const}$ for each kind of matter, one obtains
the equation of state
\begin{equation}
\label{gen-eq-st}
p=\left(\frac n3-1\right)\varepsilon.
\end{equation}
Thus, this phenomenological consideration allows one to include
most known kinds of matter characterized by their equations of
state for various $n$ \cite{Fil}. For radiation, or
ultrarelativistic gas we would have
\begin{equation}
\label{eps-4}
\varepsilon(a)=\frac{\varepsilon_4}{a^4},\quad
p=\frac\varepsilon 3.
\end{equation}
We also introduce explicitly the cosmological constant $\Lambda$,
which is associated with de Sitter vacuum with the equation of
state $p=-\varepsilon$. The Hamiltonian constraint for this model
reads
\begin{equation}
\label{constr-1}
\frac1{2a}p_a^2+\frac12a-a^3\Lambda-a^3\varepsilon(a)=0.
\end{equation}

To obtain equivalent forms of the gravitational constraint, one
could consider an arbitrary parameterization of the gauge variable,
\begin{equation}
\label{gen-param}
N=v(\tilde N,\,a),
\end{equation}
and the most general form of the constraint is
\begin{equation}
\label{gen-constr}
\frac{\partial v}{\partial\tilde N}\left(\frac1{2a}p_a^2
 +\frac12a-a^3\Lambda-a^3\varepsilon(a)\right)=0.
\end{equation}
Choosing $v(\tilde N,\,a)=\tilde N a$, one would get the
equation
\begin{equation}
\label{constr-2}
\frac12p_a^2+\frac12a^2-a^4\Lambda-a^4\varepsilon(a)=0.
\end{equation}
For a universe filled with radiation in which $\Lambda=0$, it gives
a Schr\"odinger-like equation
\begin{equation}
\label{Schro-1}
-\frac12\frac{d^2\Psi}{da^2}+\frac12a^2\Psi=E\Psi,\quad
E=\varepsilon_4.
\end{equation}
On the other hand, one can consider a universe with a non-zero
cosmological constant, but without matter. Choosing another
parameterization,
$v(\tilde N,\,a)=\displaystyle\frac{\tilde N}{a^3}$,
one would come to the equation
\begin{equation}
\label{constr-3}
\frac1{2a^4}p_a^2+\frac1{2a^2}-\Lambda-\varepsilon(a)=0.
\end{equation}
Writing down a quantum version of the constraint (\ref{constr-3}),
one faces the ordering problem. The ordering should be chosen for
the Hamiltonian operator to be Hermitian,
\begin{equation}
\label{Schro-2-1}
-\frac12\frac1{a^2}\frac d{d a}
  \left(\frac1{a^2}\frac{d\Psi}{d a}\right)
 +\frac1{2a^2}\Psi=\Lambda\Psi,
\end{equation}
or,
\begin{equation}
\label{Schro-2}
-\frac12\frac1{a^4}\frac{d^2\Psi}{da^2}
 +\frac1{a^5}\frac{d\Psi}{d a}
 +\frac1{2a^2}\Psi=\Lambda\Psi.
\end{equation}

Though the constraints (\ref{constr-1}), (\ref{constr-2}) and
(\ref{constr-3}) are completely equivalent at the classical level,
the equations (\ref{Schro-1}) and (\ref{Schro-2}) are not. Their
solutions belong to different Hilbert spaces, so that even measures
in inner products are different. In the case of Eq.\,(\ref{Schro-1})
the measure is trivial, $M(a)=1$, while in the second case, for the
Hamiltonian to be Hermitian one should choose $M(a)=a^2$. It is not
clear, if there is any relation between solutions of the equations
(\ref{Schro-1}) and (\ref{Schro-2}). Similarly, we could write down
equations for other sorts of matter. Indeed, for any
parametrization
\begin{equation}
\label{param-1}
v(\tilde N,\,a)=\tilde Na^{n-3}
\end{equation}
we would obtain a constraint
\begin{equation}
\label{constr-4}
\frac12a^{n-4}p_a^2+\frac12a^{n-2}-\varepsilon_n=0,
\end{equation}
and an appropriate Schr\"odinger equation
\begin{equation}
\label{Schro-4}
-\frac12a^{\frac n2-2}\frac d{d a}
  \left(a^{\frac n2-2}\frac{d\Psi}{d a}\right)
 +\frac12a^{n-2}\Psi=E\Psi,\quad
E=\varepsilon_n,
\end{equation}
($n=0$ for de Sitter false vacuum, $n=4$ for radiation-dominated
universe, etc.). $E$ stands for energy eigenvalues of the given
kind of matter. Let me emphasize that all these equations can be
obtained in the limits of the Dirac -- Wheeler -- DeWitt canonical
quantization scheme.

\section{Extended phase space approach}
On the contrast, in the extended phase space approach we can work with
a gauged action for pure gravitation, without matter and $\Lambda$-term
\begin{equation}
\label{action-2}
S_{(gauged)}=-\!\int\!dt\,\left[\frac12\frac{a\dot a^2}N
  -\frac12Na+\pi_0\left(\dot N-\frac{d f}{d a}\dot a\right)
  +N\dot{\bar\theta}\bar\theta\right],
\end{equation}
or, for an arbitrary parameterization of the gauge variable
$N=v(\tilde N,\,a)$,
\begin{equation}
\label{action-3}
S_{(gauged)}=-\!\int\!dt\,\left[\frac12\frac{a\dot a^2}{v(\tilde N,\,a)}
  -\frac12v(\tilde N,\,a)a
  +\pi_0\left(\dot{\tilde N}-\frac{d f}{d a}\dot a\right)
  +w(\tilde N,\,a)\dot{\bar\theta}\bar\theta\right],
\end{equation}
where a differential form of a gauge condition $\tilde N-f(a)=0$
is used, $w(\tilde N,\,a)\equiv v(\tilde N,\,a)
\displaystyle\left(\frac{\partial v}{\partial\tilde N}\right)^{-1}$,
and $\theta,\,\bar\theta$ are the Faddeev -- Popov ghosts, though a
ghost sector will not affect further consideration.

In general, a gauged action gives the gauged Einstein equations
\begin{equation}
\label{Ein.eqs}
R_{\mu}^{\nu}-\frac12\delta_{\mu}^{\nu}R
 =\kappa\left(T_{\mu(mat)}^{\nu}
 +T_{\mu(obs)}^{\nu}+T_{\mu(ghost)}^{\nu}\right),
\end{equation}
which involve, apart from energy-momentum tensor of matter fields
$T_{\mu(mat)}^{\nu}$, additional terms $T_{\mu(obs)}^{\nu}$ and
$T_{\mu(ghost)}^{\nu}$ obtained by varying the gauge-fixing and
ghost action, respectively. $T_{\mu(obs)}^{\nu}$ and
$T_{\mu(ghost)}^{\nu}$ are not true tensors, but quasi-tensors
depending on a chosen gauge conditions.

A consistent quantization procedure implies derivation of a
Schr\"odinger equation from a path integral with the gauged action
without asymptotic boundary conditions, the latter ones are thought
to be irrelevant for a closed universe \cite{SSV1,SSV2}. The
Schr\"odinger equation for a physical part of the wave function
reads
\begin{equation}
\label{gen-Schro}
\left.\left[-\frac12\sqrt{\frac{v(\tilde N,\,a)}a}
  \frac d{d a}\left(\sqrt{\frac{v(\tilde N,\,a)}a}
  \frac{d\Psi}{d a}\right)
 +\frac12 v(\tilde N,\,a)a\Psi\right]\right|_{\tilde N=f(a)}
 =E\Psi.
\end{equation}
The operator in the left-hand side of (\ref{gen-Schro}) is an
analogue of Laplacian, however, as we can see, it is
gauge-dependent,
\begin{equation}
\label{eigenvalue}
E=-\!\int\!\sqrt{-g}\,T_{0(obs)}^0\,d^3 x.
\end{equation}
For our model
\begin{equation}
\label{T-obs}
T_{\mu(obs)}^{\nu}
 ={\rm diag}\left(\varepsilon_{(obs)},\,
 -p_{(obs)},\,-p_{(obs)},\,-p_{(obs)}\right);
\end{equation}
\begin{equation}
\label{eps-obs}
\varepsilon_{(obs)}
 =\left.\frac{\dot\pi_0}{2\pi^2a^3}
   \left(\frac{\partial v}{\partial\tilde N}\right)^{-1}
   \right|_{\tilde N=f(a)};
\end{equation}
\begin{equation}
\label{gen-p-obs}
p_{(obs)}
 =\left.\varepsilon_{(obs)}\frac a{3v(\tilde N,\,a)}
   \left(\frac{\partial v}{\partial a}
   +\frac{\partial v}{\partial\tilde N}\frac{d f}{d a}
   \right)\right|_{\tilde N=f(a)}.
\end{equation}
The last formula gives the equation of state for a medium
describing by the quasi-tensor $T_{\mu(obs)}^{\nu}$ and depending
on a chosen parameterization and gauge, meanwhile
Eq.\,(\ref{gen-Schro}) determining the energy spectrum of this
medium. For the parametrization (\ref{param-1}) and the gauge
condition $\tilde N=1$ we immediately obtain the equation of state
(\ref{gen-eq-st}) from (\ref{gen-p-obs}) and Eq.\,(\ref{Schro-4})
from (\ref{gen-Schro}). Choosing in (\ref{gen-Schro})
$v(\tilde N,\,a)=\tilde N a$, we would come again to
Eq.\,(\ref{Schro-1}), while the choice
$v(\tilde N,\,a)=\displaystyle\frac{\tilde N}{a^3}$ gives
Eq.\,(\ref{Schro-2}). In other words,
Eq.\,(\ref{Schro-1}), which is believed to describe a universe
filled with radiation, corresponds to the conformal time gauge
$N=a$, and Eq.\,(\ref{Schro-2}), describing a universe with
non-zero cosmological constant, corresponds to the gauge condition
$N=\displaystyle\frac1{a^3}$. The results are completely equivalent
to those of Section 2.

Solutions to (\ref{Schro-1}), (\ref{Schro-2}) can be considered as
those of the Wheeler -- DeWitt equation for the Universe with
matter in a given state, and, at the same time, in the extended
phase space approach as solutions of a Schr\"odinger equation
associated with some gauge condition. At the level of the equations
and their solutions there is no distinctions between the two
approaches presented in Sections 2 and 3.

We can see that the medium describing by the quasi-tensor
$T_{\mu(obs)}^{\nu}$ simulates the properties of a substance
characterized by a given equation of states, in the sense that the
obtained equations are the same. Therefore, consideration of each
kind of matter is equivalent, on a phenomenological level, to a
certain choice of a gauge condition. It shows that the canonical
quantization scheme, outlined in Section 2 for arbitrary
parametrization, cannot be considered as a gauge-invariant scheme.
The condition for a gauge variable $\tilde N=1$ is used in this
scheme implicitly.

\section{Solutions for the Universe filled with two-component medium}
In the extended phase space approach we can explore even more
complicated and exotic cases. If in the limits of the approach
considered in Section 2 we can seek for eigenvalues spectrum for a
only one kind of matter in the Universe, and the presence of other
matter fields can be taken into account through additional terms in
an effective potential, the extended phase space approach allows us
to study spectra of multicomponent media. Let us turn to the
parameterization
\begin{equation}
\label{param-2}
N=v(\tilde N,\,a)=\tilde N\left(a+\frac1{a^3}\right).
\end{equation}
At small values of $a$ we have
$v(\tilde N,\,a)=\displaystyle\frac{\tilde N}{a^3}$, while at large $a$
$v(\tilde N,\,a)=\tilde N a$. The equation of state
(\ref{gen-p-obs}) under the condition $\tilde N=1$ gives
\begin{equation}
\label{p-obs}
p_{(obs)}
 =\varepsilon_{(obs)}\frac{a^4-3}{3(a^4+1)}.
\end{equation}
Again, in the limit $a\to 0$ we get
$p_{(obs)}=-\varepsilon_{(obs)}$, the equation of state for de
Sitter false vacuum, and in the limit $a\to\infty$ the equation of
state is that of radiation,
$p_{(obs)}=-\displaystyle\frac{\varepsilon_{(obs)}}3$. The
Schr\"odinger equation (\ref{gen-Schro}) for this case reads
\begin{equation}
\label{Schro-3}
-\frac12\left(1+\frac1{a^4}\right)\frac{d^2\Psi}{da^2}
 +\frac1{a^5}\frac{d\Psi}{d a}
 +\frac12\left(a^2+\frac1{a^2}\right)\Psi=E\Psi.
\end{equation}
The equations (\ref{Schro-2}) and (\ref{Schro-1}) can be obtained
as asymptotic limits of (\ref{Schro-3}) at $a\to 0$ and
$a\to\infty$, respectively. Eq.\,(\ref{Schro-3}) describes a
universe filled with a two-component medium, de Sitter vacuum
dominating at the early stage of the Universe evolution, and
radiation dominating at its later stage.

To understand the behaviour of solutions to (\ref{Schro-3}) as well
as solutions for limiting cases (\ref{Schro-2}) and (\ref{Schro-1})
one can seek for numerical solutions of these equations. A
well-known method of finding approximate eigenvalues and
eigenfunctions of a Hermitian operator consists in expanding
unknown functions onto a given basis in appropriate Hilbert space.
As was already mentioned, solutions to the equations
(\ref{Schro-2}), (\ref{Schro-1}) and (\ref{Schro-3}) belong to
different Hilbert spaces. Say, the solutions to Eq.\,(\ref{Schro-2})
exist in the Hilbert space ${\cal H}_1$ with the measure
$M(a)=a^2$, while the solutions to Eq.\,(\ref{Schro-1}) ``live'' in
the Hilbert space ${\cal H}_2$ with the measure $M(a)=1$. We should
remember that if we try to expand solutions to Eq.\,(\ref{Schro-2})
onto a basis in ${\cal H}_2$, we would obtain, in general, complex
eigenvalues, that tells about non-stability of the solutions
corresponding to a vacuum-dominated early universe and their
tendency to decay from the viewpoint of the observer located at the
later stage of the Universe evolution and using conformal time
gauge $N=a$.

We expect that in a vacuum-dominated universe (in the limit
$a\to0$) a peak of probability distribution for solutions to
Eq.\,(\ref{Schro-2}) for larger (by its absolute values)
eigenvalues tends to shift to small values of $a$. Therefore, for
larger eigenvalues of the cosmological constant $\Lambda$, which we
find solving Eq.\,(\ref{Schro-2}), the Universe appears to be
localized in the region of small $a$. On the contrary, in a
universe filled with radiation ($a\to\infty$), which is described
by solutions to Eq.\,(\ref{Schro-1}), a peak of probability
distribution for larger eigenvalues tends to shift to large values
of the scale factor. So, when the energy of this kind of matter
increases, there may be enough probability for the scale factor to
reach large values. Detailed analysis of results of numerical
calculations will be published elsewhere.

\section{Concluding remarks}
Now we return to the main question of this talk. We have
demonstrated that an equation, describing the Universe filled with
a certain sort of matter, can be obtain in the framework of the
extended phase space approach under a suitable choice of a gauge
condition. On the other hand, to investigate an energy spectrum for
this sort of matter in the Wheeler -- DeWitt quantum
geometrodynamics, one has to choose a certain form of the
gravitational constraint, i.e. a certain parameterization of a
gauge variable. Moreover, the Dirac -- Wheeler -- DeWitt
quantization procedure implies implicitly a condition on a gauge
variable $\tilde N=1$. As has been shown in
\cite{SSV1,SSV4,Shest1}, the choice of parameterization and that of
a gauge condition have a unified interpretation: together they fix
a reference frame in which we study geometry of the Universe and
matter distribution in it.

We come to the picture which is very different from ``objective
physics'' we would like to deal with. To explore some kind of matter
in a quantum universe, one has to tune up a measuring instrument in
a certain way, choosing parameterization and gauge. This changes
the form of our equations. Should we consider this fact as an
indication that our theory is wrong or accept this fact like we
accept an uncertainty principle? Let me finish on this point.


\small

\begin{thebibliography}{99}
\itemsep=-5pt
\bibitem{BK}
 J.D. Brown and K.V. Kucha\v r,
 {\it Phys. Rev.\/} {\bf D 51}, 5600 (1995).
\bibitem{MM}
 S. Mercuri and G. Montani,
 {\it Int. J. Mod. Phys.\/} {\bf D 13}, 165 (2004).
\bibitem{BM}
 M.V. Battisti and G. Montani.
 ``Evolutionary Quantum Dynamics of a Generic Universe'',
 gr-qc/0604049.
\bibitem{SS1}
 T.P. Shestakova and C. Simeone,
 {\it Grav. \& Cosmol.\/} {\bf 10}, 161 (2004).
\bibitem{SS2}
 T.P. Shestakova and C. Simeone,
 {\it Grav. \& Cosmol.\/} {\bf 10}, 257 (2004).
\bibitem{SSV1}
 V.A. Savchenko, T.P. Shestakova and G.M. Vereshkov,
 {\it Int. J. Mod. Phys.\/} {\bf A 14}, 4473 (1999).
\bibitem{SSV2}
 V.A. Savchenko, T.P. Shestakova and G.M. Vereshkov,
 {\it Int. J. Mod. Phys.\/} {\bf A 15}, 3207 (2000).
\bibitem{SSV3}
 V.A. Savchenko, T.P. Shestakova and G.M. Vereshkov,
 {\it Grav. \& Cosmol.\/} {\bf 7}, 18 (2001).
\bibitem{SSV4}
 V.A. Savchenko, T.P. Shestakova and G.M. Vereshkov,
 {\it Grav. \& Cosmol.\/} {\bf 7}, 102 (2001).
\bibitem{Fil}
 M.L. Fil'chenkov,
 {\it Phys. Lett.\/} {\bf B 441}, 34 (1998).
\bibitem{DF}
 I.G. Dymnikova and M.L. Fil'chenkov,
 {\it Phys. Lett.\/} {\bf B 545}, 214 (2002).
\bibitem{Shest1}
 T.P. Shestakova,
 {\it in:\/} Proceedings of the IV International
 Conference ``Cosmion-99'',
 {\it Grav. \& Cosmol.\/} {\bf 6}, Supplement, 47 (2000).
\end{thebibliography}
\end{document}